\documentclass[usegraphicx]{mn2e}

\usepackage{amssymb}
\usepackage{mathptmx}
\usepackage{url}
\usepackage{amssymb}

\newcommand{\Mpc}{\rm\thinspace Mpc}

\newcommand{\km}{\rm\thinspace km}

\newcommand{\cm}{\rm\thinspace cm}
\newcommand{\s}{\rm\thinspace s}

\newcommand{\Msun}{\hbox{$\rm\thinspace M_{\odot}$}}

\newcommand{\kmps}{\hbox{$\km\s^{-1}\,$}}

\newcommand{\kmpspMpc}{\hbox{$\kmps\Mpc^{-1}$}}

\newcommand{\pcmsq}{\hbox{$\cm^{-2}\,$}}

\begin{document}

\title{An X-ray absorption analysis of the high-velocity system in NGC 1275}
\author[K. Gillmon, J.S. Sanders and A.C. Fabian]
{K. Gillmon${}^{1,2}$, J.S. Sanders${}^1$ and A.C. Fabian${}^1$\\
${}^1$Institute of Astronomy, Madingley Road, Cambridge. CB3 0HA\\
${}^2$Center for Astrophysics and Space Astronomy, Department of Astrophysical and Planetary Sciences, University of Colorado, Boulder, CO 80309-0389, U.S.A.}
\maketitle

\begin{abstract}
We present an X-ray absorption analysis of the high-velocity system
(HVS) in NGC 1275 using results from a deep 200 ks \emph{Chandra}
observation.  We are able to describe the morphology of the HVS in
more detail than ever before.  We present an HST image for comparison,
and note close correspondence between the deepest X-ray absorption and
the optical absorption.  A column density map of the HVS shows an
average column density $N_\mathrm{H}$ of $1 \times 10^{21}\pcmsq$ with
a range from $\sim5 \times 10^{20}$ to $5 \times 10^{21}\pcmsq$.  From the
$N_\mathrm{H}$ map we calculate a total mass for the absorbing gas in
the HVS of $(1.32 \pm 0.05) \times 10^{9} \Msun$ at solar abundance.
75 per cent of the absorbing mass is contained in the four regions of
deepest absorption.  We examine temperature maps produced by spectral
fitting and find no direct evidence for shocked gas in the HVS.  Using
deprojection methods and the depth of the observed absorption, we are
able to put a lower limit on the distance of the HVS from the nucleus
of 57 kpc, showing that the HVS is quite separate from the body of
NGC\,1275.
\end{abstract}

\begin{keywords}  
  galaxies: clusters: individual: Abell 426 --- galaxies: individual:
  NGC 1275 --- galaxies: interactions --- X-rays: galaxies
\end{keywords}

\section{Introduction}
Optical imaging and spectroscopy first established the existence of
two distinct emission line systems toward NGC 1275, the central cD
galaxy of the Perseus cluster: a low-velocity component associated
with the cD galaxy itself at 5200 \kmps and a high-velocity component
at 8200 \kmps projected nearby on the sky (Minkowski 1955,1957).
Since then, observations in H$\alpha$, radio, optical, and X-ray bands
have revealed a web of H$\alpha$ filaments surrounding and comoving
with the cD galaxy (Lynds 1970; Heckman et al 1989) as well as
absorption by the high-velocity system placing it in front of the
cluster core (De Young, Roberts, $\&$ Saslaw 1973; Rubin et al 1977;
Kent $\&$ Sargent 1979; Boroson 1990; Fabian et al 2000).  The
high-velocity system (HVS) is likely a galaxy falling into the Perseus
cluster at 3000 \kmps; however, the complexity of the NGC 1275 system
has left many unanswered questions.  In particular, the nature of the
interaction between the infalling galaxy, the cD galaxy, and the
H$\alpha$ filaments is not well understood.

Early models of NGC 1275 describe two galaxies in the process of
merging (Minkowski 1955, 1957).  Support for this theory includes
possible spatial correspondence between the low and high-velocity
systems (Hu et al 1983; Unger et al 1990), gas at intermediate
velocities between 5200 and 8200 \kmps (Ferruit et al 1997), and
possible disruption and associated star formation in both the HVS and
cD galaxy (Hu et al 1983; Unger et al 1990; Boroson 1990; Shields $\&$
Filippenko 1990).  However, all of these phenomena could potentially
be explained by a previous interaction of the low and/or high-velocity
system with a third gas-rich galaxy or system of galaxies (Holtzman et
al 1992; Conselice, Gallagher, $\&$ Wyse 2001), or by influences from
the surrounding dense intracluster medium (ICM) (Fabian $\&$ Nulsen
1977; Sarazin 1988; Boroson 1990; Caulet et al 1992).

The mass of the HVS is an important parameter that could help classify
the galaxy as well as determine its history.  Van Gorkom $\&$ Ekers
(1983) find an upper limit of $2.5\times 10^9\Msun$ on the HI mass of
the galaxy that classifies it as type sC or earlier; but it has also
been suggested that low mass content could be evidence of ram pressure
stripping or a past interaction with another galaxy (Conselice et al
2001).  Also, evidence of shocked gas would support the theory that
the HVS is undergoing current interaction.

In addition to the mass of the HVS, its position relative to the
cluster core still needs to be determined.  Although it is accepted
that the HVS lies in front of the cluster core, attempts to quantify
the distance have resulted in various inconclusive estimates (van
Gorkom $\&$ Ekers 1983; Pedlar et al 1990; Fabian et al 2000).
Because the NGC 1275 system consists of multiple, possibly
interrelated components, a more accurate determination of the mass and
position of the HVS would aid characterization of the entire system.

The hot ICM provides a background of X-rays for viewing the HVS in
X-ray absorption.  A previous 25 ks \emph{Chandra} observation (Fabian
et al 2000) yielded estimates for the average column density and total
mass of the HVS.  We now present results from a deep 200 ks
\emph{Chandra} observation which provides a more lucid picture of the
HVS morphology and allows both column density and total mass to be
more accurately determined.  In addition, we use temperature mapping
to search for shocked gas, and deprojection methods to place a lower
limit on the distance of the HVS from the cluster core.

The Perseus cluster is at a redshift of 0.0183.  We assume that $H_0 =
70 \kmpspMpc$, which gives a luminosity distance to the cluster of 80
Mpc and a scale of 1 kpc corresponds to about 2.7 arcsec.

\section{Morphology}
Fig. 1 shows a 1.77 by 0.89 arcmin section of the 200 ks X-ray image
covering the energy range 0.5-1 keV.  The high-velocity system can
clearly be seen in absorption to the north of the bright NGC 1275
nucleus which is at RA 3 19 48, Dec +41 30 42 J2000.  Projected behind
the HVS are regions of low X-ray luminosity and surrounding bright
rims associated with the radio lobes of the central source (Fabian et
al 2000).  At least 4 areas of deep absorption are clearly visible
(one just to the east of the nucleus, two just to the west of the
nucleus, and one further west at RA 45 sec).

For comparison, an HST image of the same region (filter 702W, dataset
u2pf0404b) is shown in Fig. 2 (the image has been unsharp masked in
\textsc{gimp}\footnote{http://www.gimp.org/} to bring out the
absorption features).  We note close correspondence between the deep
X-ray absorption features and the optical absorption features.

Also, the point source at RA 03 19 46.35 Dec +41 30 43.6 detected in
the X-ray image (Fig. 1) is visible in the HST image, suggesting it
may be associated with star clusters in the HVS.  The point source at
RA 03 19 48.06 Dec +41 31 01.9 detected in the X-ray image (Fig. 1) is
not visible in the HST image.

In addition to the regions of deep absorption, it is likely that
weaker, more diffuse absorption is occurring between and around the
deep absorption.  However, it is difficult to discern this diffuse
absorption in Fig. 1 due to the change in background brightness over
the HVS.  Fig. 3 shows the ratio of a soft (0.5-1 keV) and a hard (2-5
keV) X-ray image binned to a resolution of 1.96 arcsec (4 pixels).
The regions of likely diffuse absorption are indicated a-d.  These
regions are marked by intermediate ratios between the lowest ratios,
occurring at the positions of deep absorption, (dark) and the
characteristic ratios everywhere else (light).  Since it appears
unlikely that variations of the ICM temperature, abundance, or
galactic absorption could cause these features, we are confident that
(a-d) are in fact regions of absorber in the HVS.

\section{X-ray Spectral Analysis}
In order to quantify the absorption column density structure of the
HVS, we fit spectra extracted from bins of several different spatial
resolutions.  First, the region seen in all previous figures was
divided into a grid of 7.84 arcsec (16 pixel) square bins.  Spectra
were extracted from each of these bins and fit from 0.6-8 keV (this
energy range assumed from hereon for all spectral fits) with a single
temperature \textsc{mekal} spectral model (Mewe, Gronenschild $\&$ van
den Oord 1985; Liedahl, Osterheld $\&$ Goldstein 1995) absorbed by a
\textsc{phabs} model (Balucinska-Church $\&$ McCammon 1992).  Since
the HVS lies at relatively low $z$, the galactic and HVS absorption
were both modelled by the single \textsc{phabs} component.  The
absorption column, temperature, and \textsc{mekal} abundance (assuming
solar ratios of Anders $\&$ Grevesse 1989) were left free for these
fits. The mean reduced-$\chi^2$ of the fits was 1.1 and no significant
residuals were seen at low energy. The image was then divided into a
higher resolution grid of 1.96 arcsec (4 pixel) square bins.  Spectra
were again extracted and fit with a single temperature \textsc{mekal}
model and \textsc{phabs} absorber, but this time the temperature and
\textsc{mekal} abundance were fixed at the appropriate value obtained
by fitting the low-resolution grid, thus leaving only the absorption
column density and normalization as free parameters. The mean
reduced-$\chi^2$ of the fits was 1.0.  The rational behind this
two-scale approach is that we expect the background temperature and
abundance to vary on smaller scales than the absorber. The result of
these fits yielded a map of the combined galactic and HVS
$N_\mathrm{H}$.  In order to isolate the HVS $N_\mathrm{H}$, we took
the mean of a 30 by 50 arcsec region to the left of the HVS to be an
indication of the galactic $N_\mathrm{H}$ contributing to each bin,
and subtracted this value from every bin over the entire region.  The
final results of the 1.96 arcsec resolution fits, with galactic
$N_\mathrm{H}$ subtracted, are seen as an HVS column density map in
Fig. 4.  The average $N_\mathrm{H}$ of the HVS is about $1 \times
10^{21}\pcmsq$ with a range from $5 \times 10^{20}$ to $5 \times
10^{21}\pcmsq$.  The 1$\sigma$ uncertainty in $N_\mathrm{H}$ is $0.3
\times 10^{21}\pcmsq$ for the lowest values, and $0.6 \times
10^{21}\pcmsq$ for the highest values.  Our previous conclusion that
diffuse absorption is occurring between and around the regions of deep
absorption is supported by the correspondence between Fig. 3 and Fig.
4. Summing the columns of gas leads to a total absorbing mass of
$(1.32 \pm 0.05) \times 10^9 \Msun$ at solar abundance.  75 per cent
of the absorbing mass is contained in the four regions of deepest
absorption.

We also fit spectra from a 0.49 arcsec square grid, minimizing $C$
statistics (Cash 1979) rather than $\chi^2$ due to the low number of
counts per spectral bin.  Fig. 5 shows the resulting column density
map (including galactic and HVS), smoothed with a Gaussian of 1 pixel.
Though the errors on these column densities are large, the smoothed
image provides a qualitative illustration of the HVS column density
structure.

The \emph{Chandra} datasets used here are those presented in Fabian et
al (2003), corrected for time-dependent gain shift using the
\textsc{corr\_tgain} utility (Vikhlinin 2003) and the
corrgain2002-05-01.fits correction file.  In all cases, a background
spectrum generated from blank-sky observation was used, however the
contribution of this background was insignificant.  Response matrices
and auxiliary responses based on the centre of the cluster were
applied over the entire 1.77 by 0.89 arcmin region, and the auxiliary
response files were corrected for low energy QE
degradation\footnote{http://cxc.harvard.edu/cal/Links/Acis/acis/Cal\_prods/qeDeg/}
using the \textsc{corrarf} routine (Vikhlinin 2002) to apply the
\textsc{acisabs} absorption correction (Chartas \& Getman 2002).  We
found that using a \textsc{contamarf}
correction\footnote{http://space.mit.edu/CXC/analysis/ACIS\_Contam/script.html}
instead of a \textsc{corrarf} correction caused the $N_\mathrm{H}$
values across the entire region to increase by $0.2 \times
10^{21}\pcmsq$.  However, this difference can be incorporated into the
subtraction of the galactic $N_\mathrm{H}$, thus the choice of
correction does not affect the HVS $N_\mathrm{H}$.  In the 7.84 arcsec
and 1.96 arcsec fits, the spectra were binned to include a minimum of
20 counts per spectral bin.  The spectra from the 0.49 arcsec grid
were not binned due to low numbers of counts.  On all column density
maps, the circular region of radius about 5 arcsec centred on the NGC
1275 nucleus should be ignored.

By examining column density maps at several different spatial
resolutions, we suspect that the $N_\mathrm{H}$ of the HVS is likely
varying on scales smaller than that resolved by the map at 1.96 arcsec
resolution.  If the bin from which the spectrum is extracted contains
multiple absorption column density components, then the contribution
to the spectrum from the less absorbed areas will dominate, causing
the column density obtained from spectral fitting to be lower than the
mean column density weighted by the covering fraction of each
component.  To quantify this effect we faked a spectrum with 2
\textsc{mekal} components of equal normalization, both with a
temperature of 3 keV and an abundance of 0.6 solar.  One of the
components was \textsc{phabs} absorbed by a hydrogen column density of
$10^{21}\pcmsq$ and the other by a hydrogen column density of
$10^{22}\pcmsq$.  We fit the faked spectrum with a single temperature
\textsc{mekal} spectral model and \textsc{phabs} absorber.  The best
fit column density was $2 \times 10^{21}\pcmsq$ within uncertainties,
independent of S/N, rather than the mean of $5.5 \times
10^{21}\pcmsq$.  As a result of this effect, the absorbing mass
calculated from the 1.96 arcsec column density map is likely to be a
lower limit on the true value.  A further systematic uncertainty
derives from the abundance of the gas. The main absorber in the HVS is
likely to be oxygen and $N_\mathrm{H}$ and the total absorbing mass
both depend on the HVS metal abundance. The \textsc{phabs} model
assumes solar abundance to calculate $N_\mathrm{H}$, and we assumed
solar abundance to calculate the total absorbing mass.  Therefore, if
the metal abundance is less than solar, the $N_\mathrm{H}$ and total
absorbing mass will be greater than that found above, and vice versa
if greater than solar.

In order to search for high temperature shocked gas, we also fit the
7.84 arcsec resolution grid with 2 \textsc{mekal} temperature
components.  The \textsc{mekal} abundances were left free and tied,
and a single \textsc{phabs} absorber was applied.  A comparison of the
two temperature maps obtained from these fits shows no indication of
high temperature gas associated spatially with the HVS.  Based on this
fairly simple approach, we see no direct evidence for shocked gas in
the HVS. For a more quantitative approach we have set the upper
temperature in these two-temperature spectral fits to the likely shock
temperature of 11~keV. We then find the emission integral of hot gas
$n^2 \ell,$ where $\ell$ is the length of the shock region along the
line of sight, to be less than about $3\times 10^{20}$~cm$^{-5}$. If
$\ell$ is comparable to the size of the HVS of about 10~kpc and the
density jump is the strong shock value of 4, then the preshock
intracluster medium must then have a density less than about
0.025~cm$^{-3}$. From the density profile given in Fabian et al
(2003), this corresponds to a radius greater than 25~kpc from the
nucleus of NGC\,1275.

\section{Distance of HVS from the nucleus of NGC\,1275}
We have above given an estimate of the minimum distance of the HVS from
NGC\,1275 based on the lack of high temperature emission from shocked
gas. We now obtain a tighter constraint from the depth of the
absorption seen. The closer that the HVS is to the nucleus the more
the absorption will be filled in by emission from intervening cluster gas.

We measured an average count rate of 0.25 counts/pixel (each pixel
0.49 arcsec square) in the 0.4-0.7 keV band over the 5 by 5 pixel
region of deepest absorption.  We also note that there are regions of
3 by 3 pixels with zero counts.  Since some or all of these counts may
be transmitted from behind the HVS, we use the rate of 0.25
counts/pixel to place a lower limit on the distance of the HVS from
the cluster core, assumed here to be the point source 3C 84.

Using a region from 120 deg to 60 deg measured clockwise from the
east-west line through the cluster core, we created a surface
brightness profile, $S$($R$), of the ICM emission where $R$ is the
projected radial distance from the cluster core in the plane of the
sky.  We then fit the profile, for $R>160$ arcsec, with a power law.
Assuming the emissivity, $\epsilon$ ($r$) where $r$ is the radial
distance from the cluster core, is also a power law and that the
surface brightness and emissivity are related by

\begin{equation} S(R)=\int^{\infty}_{-\infty} \epsilon(r)\,\mathrm{d}\ell \,,\end{equation}
where $\ell$ is the position along the line of sight and \( r=(R^2+\ell^2)^{1/2} \),
we determined $\epsilon$ ($r$) from (1)$:$

\begin{equation} \epsilon(r)=25365\,\,r^{-3.14}\,\,\,\,\,\mbox{counts}\,\,\mbox{pixel}^{-2}\,\mbox{arcsec}^{-1} \end{equation}
Let the distance from the HVS to the cluster core be L.  Assuming
$R\ll$ L, then $\epsilon$ ($r$) $\approx$ $\epsilon(\ell)$.  We know
that the integral of $\epsilon(\ell)$ with respect to $\ell$ from L to
infinity cannot exceed 0.25 counts/pixel.  Solving the inequality for
L yields a lower limit on the distance of the HVS from the cluster
core of 153 arcsec or 57 kpc.

This limit would be complicated if the HVS is falling {\it exactly}
along our line of sight so we are viewing it along a wake of shocked
gas. A detailed numerical computation is required to model this
situation but we doubt that it would make a large difference to the
above result.

The column density in the hot gas lying from 60 to 10~kpc is about
$5\times 10^{21}$~cm$^{-21}$. This is similar to the highest column
densities in the HVS, so most of the gas will be stripped before
reaching NGC\,1275.

\section{Summary and Conclusions}
We note close correspondence between the morphology of the HVS
depicted in X-ray absorption, optical absorption, and column density
maps.  The most dominant features of the HVS are 4 high column density
regions of deep absorption surrounded by patches of more diffuse
material.  By fitting spectra extracted from bins of several different
spatial resolutions, we were able to obtain a 1.96 arcsec resolution
map of the HVS column density.  From this map we calculate a total
absorbing mass for the gas in the HVS of at least $(1.32 \pm 0.05)
\times 10^{9} \Msun$ at solar abundance. This is below the HI upper limit of
van Gorkom and Ekers (1983).  We have analyzed temperature maps also
produced from spectral fitting and found no direct evidence for
shocked gas in the HVS.  Finally, using the average count rate in the
regions of deepest absorption, we were able to put a lower limit on
the distance of the HVS from the cluster core of 57 kpc.  The mass
in combination with the lack of obvious shocked gas and the
distance of the HVS from the cluster core support the theory that the
low and high-velocity systems are not interacting.  Since the HVS
appears to be falling into the Perseus cluster at 3000 \kmps, any
merger or collision with NGC 1275 is still in its future.

\begin{figure*}
 \centering
  \includegraphics[width=0.99\textwidth]{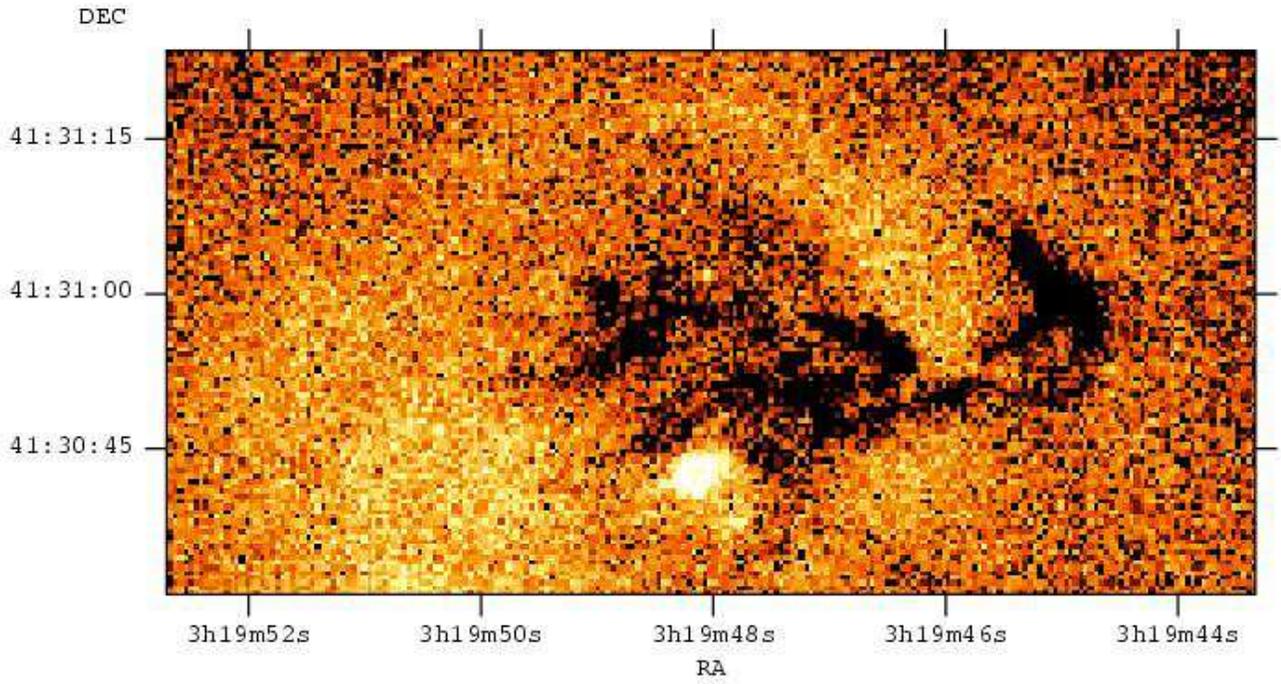}
  \caption{Central region of NGC 1275 in the 0.3-0.8 keV band.  Pixels are 0.49 arcsec in dimension and the entire image is 1.77 by 0.89 arcmin.  North is to the top and east is to the left in this image.  The high-velocity system is seen in absorption to the north of the bright nucleus at RA 3 19 48, Dec +41 30 42.}  
  \label{Fig. 1}
\end{figure*}

\begin{figure*}
  \centering
  \includegraphics[width=0.99\textwidth]{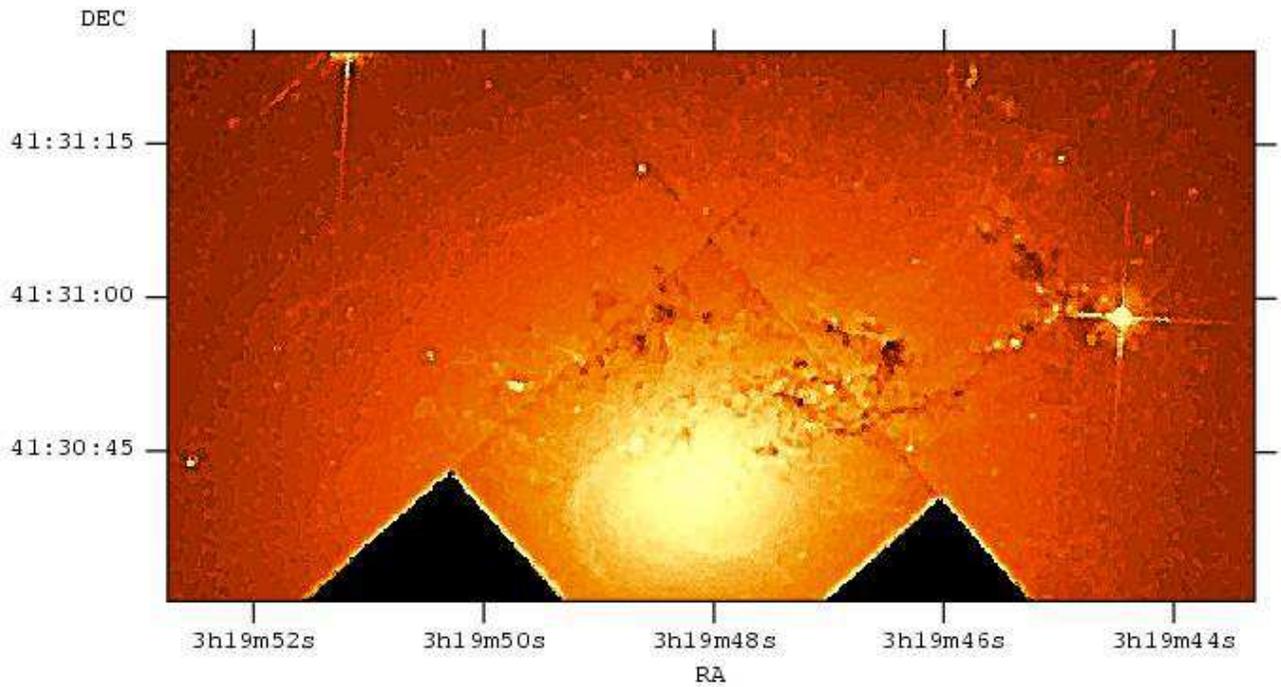}
  \caption{HST image (filter F702W) of the same region shown in Fig.
  1-4.  Close correspondence can be seen between the deepest X-ray
  absorption features and the optical absorption features.}
  \label{Fig. 2}
\end{figure*}

\begin{figure*}
 \centering
 \includegraphics[width=0.99\textwidth]{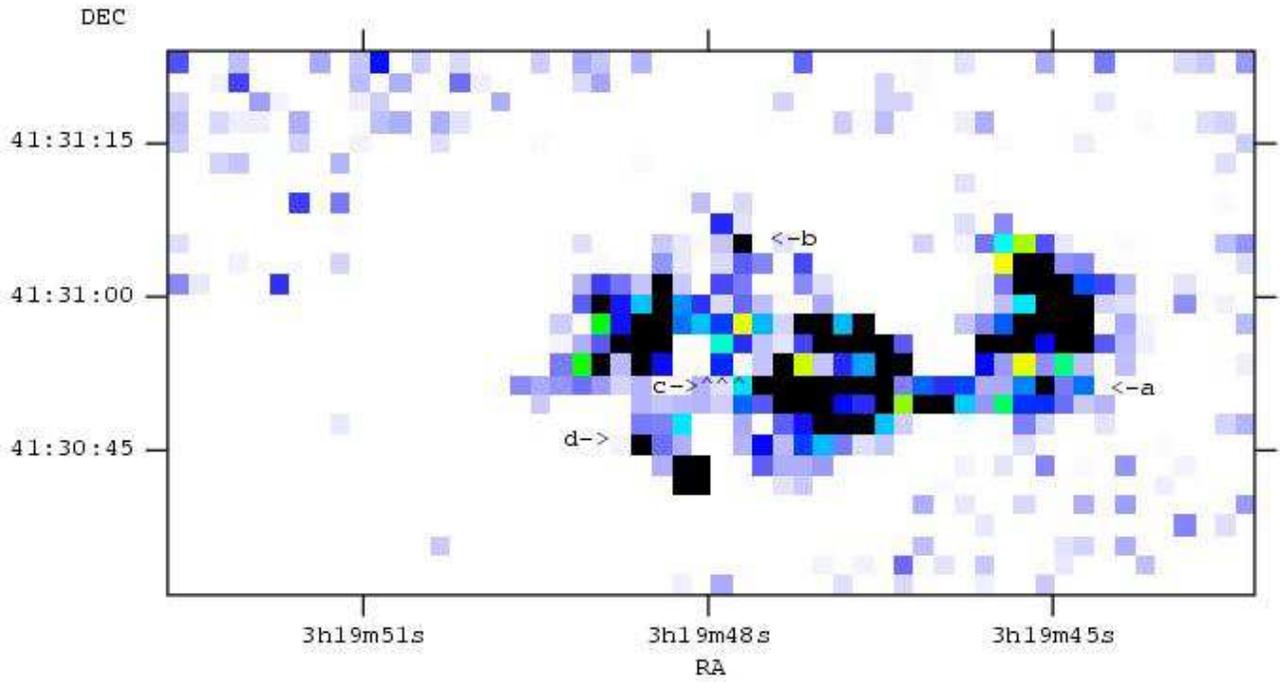}
 \caption{Ratio of a soft (0.5-1 keV) and a hard (2-5 keV) X-ray image binned to a resolution of 1.96 arcsec.  Regions of diffuse absorption are indicated a-d.} 
 \label{Fig. 3}
\end{figure*}

\begin{figure*}
  \centering
  \includegraphics[width=0.99\textwidth]{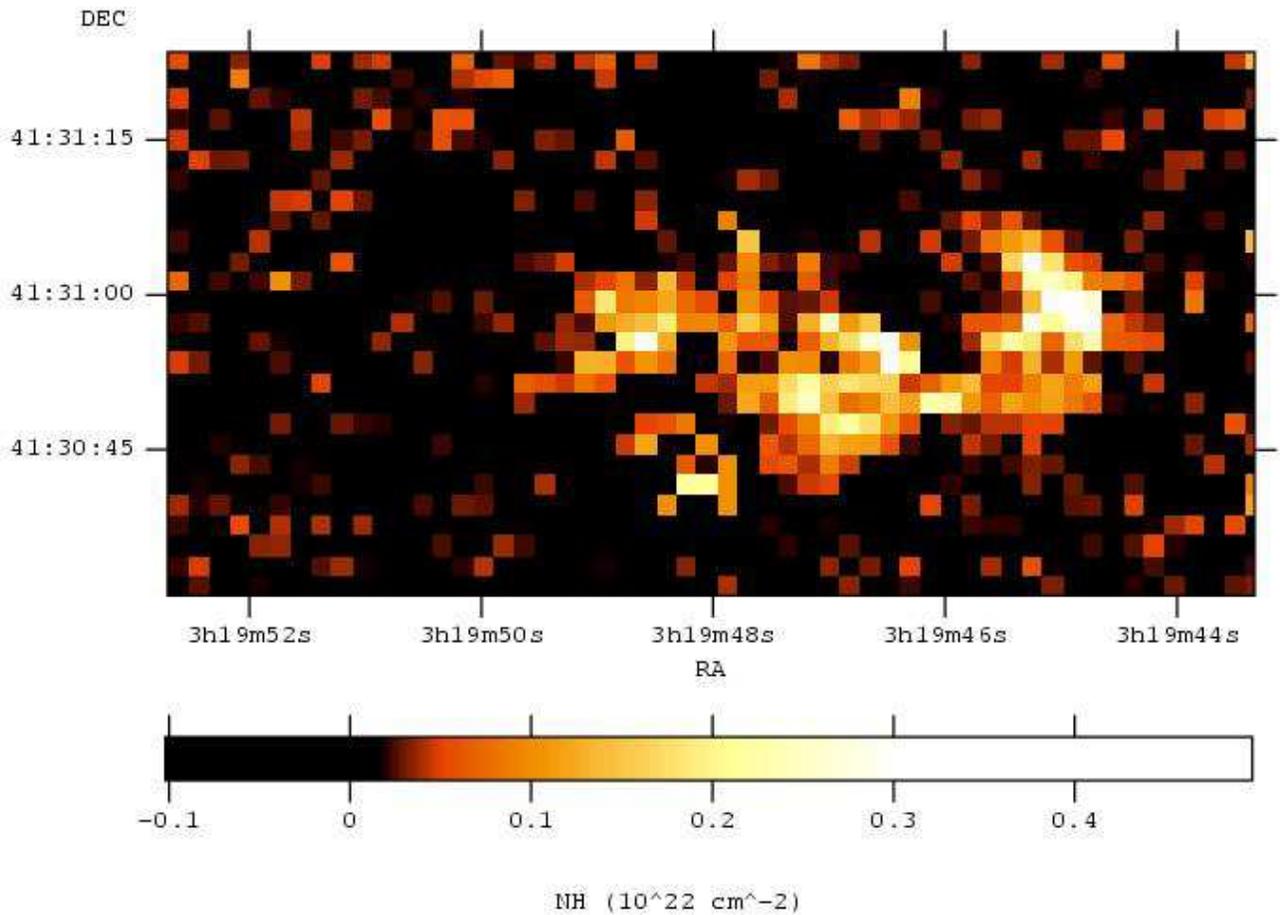}
  \caption{1.96 arcsec resolution map of HVS $N_\mathrm{H}$ (an estimate for the mean galactic $N_\mathrm{H}$ per bin has been subtracted from each bin).}
  \label{Fig. 4}
\end{figure*}

\begin{figure*}
  \centering
  \includegraphics[width=0.99\textwidth]{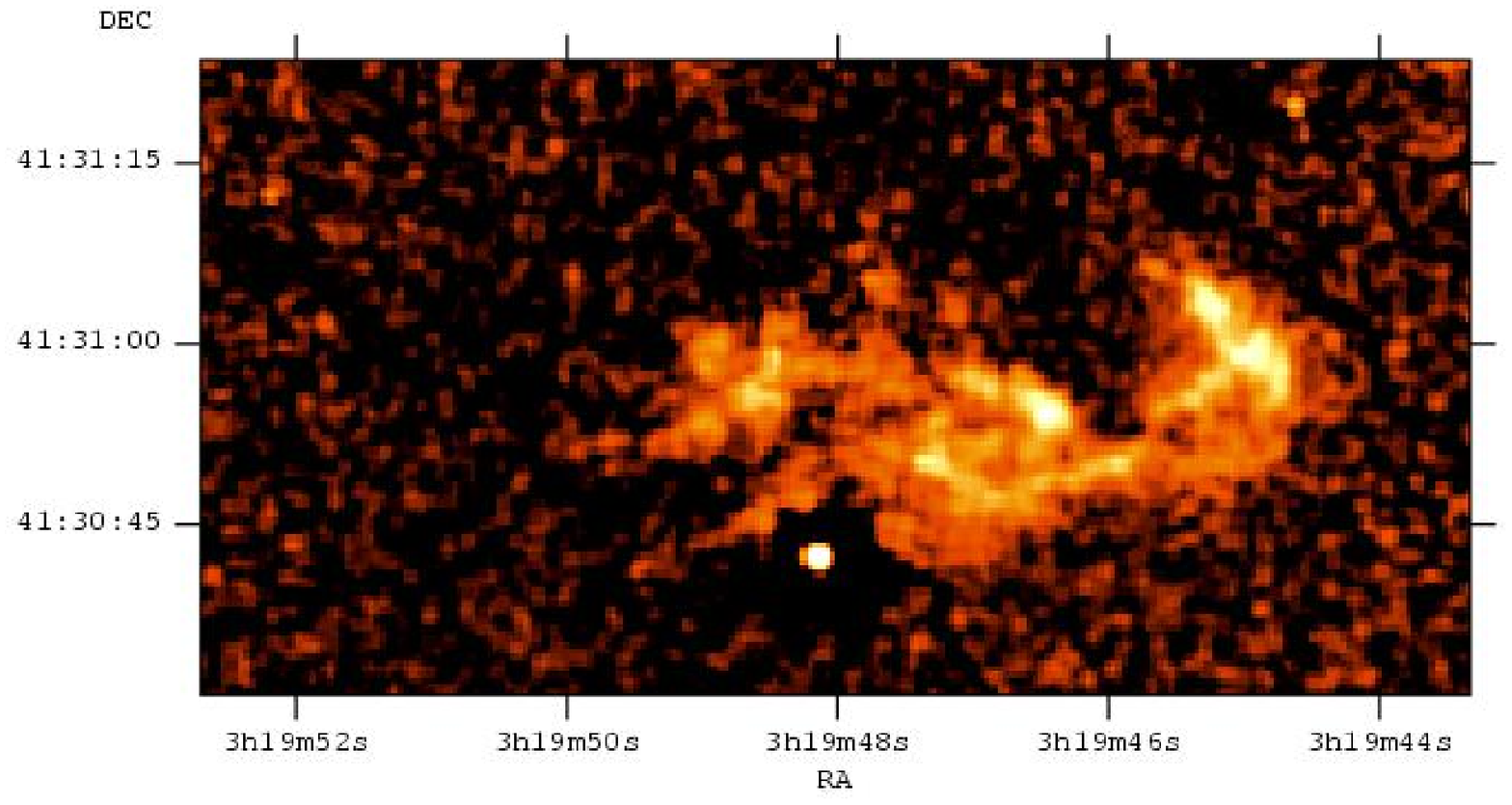}
  \caption{0.49 arcsec resolution map of total $N_\mathrm{H}$ (including galactic and HVS), smoothed with a Gaussian of 1 pixel.}
  \label{Fig. 5}
\end{figure*}

\section*{Acknowledgements}
KG thanks the Boettcher Foundation for their support, and Adrian
Turner, Lisa Voigt, and Matthew Worsley for all of their help.  ACF
thanks the Royal Society for their support.

The \emph{HST} data presented in this paper were obtained from the
Multimission Archive at the Space Telescope Science Institute (MAST).
STScI is operated by the Association of Universities for Research in
Astronomy, Inc., under NASA contract NAS5-26555.


\begin{thebibliography}{}
\bibitem{} Anders E., Grevesse N., 1989, Geochimica et Cosmochimica Acta, 53, 197
\bibitem{} Balucinska-Church M., McCammon D., 1992, ApJ, 400, 699
\bibitem{} Boroson T.A., 1990, ApJ, 360, 465
\bibitem{} Cash W., 1979, ApJ, 228, 939
\bibitem{} Caulet A., Woodgate B.E., Brown L.W., Gull T.R., Hintzen P., Lowenthal J.D., Oliversen R.J., Ziegler M.M., 1992, ApJ, 388, 301
\bibitem{} Chartas G., Getman K., 2002, http://www.astro.psu.edu/\-users/\-chartas/\-xcont\-dir/\-xcont\-.html
\bibitem{} Conselice C.J., Gallagher J.S., Wyse R.F.G., 2001, AJ, 122, 2281
\bibitem{} De Young D.S., Roberts M.S., Saslaw W.C., 1973, ApJ, 185, 809
\bibitem{} Fabian A.C., Nulsen P.E.J., 1977, MNRAS, 180, 479
\bibitem{} Fabian A.C. et al., 2000, MNRAS, 318, L65
\bibitem{} Fabian A.C., Sanders J.S., Allen S.W., Crawford C.S., Iwasawa K., Johnstone R.M., Schmidt R.W., Taylor G.B., 2003, MNRAS, in press
\bibitem{} Ferruit P., Adam G., Binette L., Pecontal E., 1997, NewA, 2, 345
\bibitem{} Heckman T.M., Baum S.A., van Breugel W.J.M., McCarthy P., 1989, ApJ, 338, 48
\bibitem{} Holtzman J.A. et al., 1992, AJ, 103, 691
\bibitem{} Hu E.M., Cowie L.I., Kaaret P., Jenkins E.B., York D.G., Roesler F.L., 1983, ApJ, 275, L27
\bibitem{} Kent, S.M., Sargent W.L.W., 1979, 230, 667
\bibitem{} Liedahl D.A., Osterheld A.L., Goldstein W.H., 1995, ApJ, 438, L115
\bibitem{} Lynds C.R., 1970, ApJ, 159, L151
\bibitem{} Mewe R., Gronenschild E.H.B.M., van den Oord G.H.J., 1985, A\&AS, 62, 197
\bibitem{} Minkowski R., 1955, Carnegie Yearbook, 54, 25
\bibitem{} Minkowski R., 1957, IAU Sympn 4, p107
\bibitem{} Pedlar A., Ghataure H.S., Davies R.D., Harrison B.A., Perley R., Crane P.C., Unger S.W., 1990, MNRAS, 246, 477
\bibitem{} Rubin V.C., Ford K., Peterson C.J., Oort J.H., 1977, ApJ, 211, 693
\bibitem{} Sarazin C.L., 1988, X-Ray Emissions from Clusters of Galaxies (Cambridge: Cambridge Univ. Press)
\bibitem{} Shields J.C., Filippenko A.V., 1990, ApJ, 353, L7
\bibitem{} Unger S.W., Taylor K., Pedlar A., Ghataure H.S., Penston M.V., Robinson A., 1990, MNRAS, 242, 33P
\bibitem{} van Gorkom J.H., Ekers R.D., 1983, ApJ, 267, 528
\bibitem{} Vikhlinin A., 2002, http://hea-www.harvard.edu/$\sim$alexey/corrarf/
\bibitem{} Vikhlinin A., 2003, http://\-asc.harvard.edu/\-cont-soft/\-software/\-corr\-\_tgain.\-1.0.\-html


\end{thebibliography}
\end{document}